# Linear and nonlinear optical tuning with $Sb_2S_3$-based metasurfaces

Amin Zamani,[†] Meibao Qin,[*,‡] Gabriel Sanderson,[†] Lu Zhang,[¶] Sara Moujdi,[†] Qiwei Miao,[¶] Ze Zheng,[†] Mohammadhossein Momtazpour,[†] Christopher J. Mellor,[§] Wending Zhang,[¶] Ting Mei,[¶] Zakaria Mansouri,[†] Lei Xu,[*,†] and Mohsen Rahmani[†]

† *Department of Engineering, School of Science and Technology, Nottingham Trent University, NG11 8NS, Nottingham, United Kingdom*

‡ *School of Education, Nanchang Institute of Science and Technology, China*

¶*Key Laboratory of Light Field Manipulation and Information Acquisition, Ministry of Industry and Information Technology, School of Physical Science and Technology, Northwestern Polytechnical University, 710129, Xi'an, China*

§*School of Physics and Astronomy, Faculty of Science, University of Nottingham, NG7 2RD, Nottingham, United Kingdom*

E-mail: qinmb@ncpu.edu.cn; lei.xu@ntu.ac.uk

**Abstract**

In the past decade, phase-change metasurfaces have attracted significant attention for reconfigurable photonic devices owing to their ability to tune optical responses through controllable intermediate crystallisation states between amorphous and crystalline phases. Despite their potential to revolutionise applications in optical switching, communications, and signal processing, a major challenge in phase-change metasurfaces is achieving large optical modulation while maintaining low optical loss and strong resonant field confinement. Here, we experimentally demonstrate the unique properties of $Sb_2S_3$ metasurfaces on monolithic and hybridised $Sb_2S_3$-Si platforms. Their advantages are particularly pronounced in the telecommunication spectral range, where both phases remain nearly lossless together with a large and stable refractive index contrast compared to other counterparts. The monolithic metasurface enables transmission modulation depths of up to 92% and resonance shifts as large as ∼150 nm at the telecommunication wavelengths through laser-induced phase transitions from amorphous to ∼ 50% polycrystalline states. Furthermore, we demonstrate that integrating a silicon overlayer introduces high-Q hybrid resonances with enhanced near-field confinement, enabling comparable modulation, experimentally, with nearly half the laser switching power required as compared to the monolithic structure. Beyond linear optical switching, the broad resonance tunability of the $Sb_2S_3$ metasurface, together with the strong third-order nonlinearity of these materials, provides a promising platform for tunable nonlinear light generation. By exciting the $Sb_2S_3$ metasurfaces with a fixed broadband pump, we experimentally demonstrate tunable third-harmonic generation emission over a ∼40 nm spectral range through phase-change transition. Such tunablity provides a versatile route for integrating $Sb_2S_3$ with intrinsically highly nonlinear materials to enable the combination of large optical tunability and efficient nonlinear light generation. We have proven this concept via our hybrid $Sb_2S_3$-Si platform, which shows a threefold enhancement in tunable THG emission arising from the combined nonlinear responses of $Sb_2S_3$ and silicon, and benefiting from enhanced electromagnetic field confinement enabled by the metasurface's high-Q resonances. These results highlight the potential of $Sb_2S_3$ monolithic and hybrid metasurfaces for compact low-power optical switching, tunable nonlinear emitters, and visible light sources.

# Introduction

Dielectric metasurfaces offer a powerful platform for controlling light–matter interactions at the nanoscale, enabling compact and efficient manipulation of the phase, amplitude, polarisation, and propagation of light.[1–5] Their ability to confine light within the volume of each meta-atom, combined with intrinsically low optical absorption, makes them particularly well-suited for a wide range of linear and nonlinear optical functionalities, including beam steering, optical switching, nonlinear frequency conversion and nonlinear wavefront control.[6–11] Despite these advances, most existing metasurface devices remain static after fabrication. To overcome these limitations, tunability in metasurfaces has been explored through various mechanisms, including mechanical,[12] thermal,[13–15] electro-optic,[16] and laser-induced methods.[17] However, conventional dielectric metasurface-based optical tunings generally face a trade-off between achieving a high modulation depth and maintaining low optical loss in the telecommunication band, while large-scale spectral tuning remains challenging in conventional photonic materials. Among these approaches, laser tuning of phase-change materials (PCMs) provides a particularly attractive approach to realise large-scale dynamic light control due to their large refractive index modulation, low optical losses and nonvolatile switching capability.[18–22] Optical tuning and switching have been widely demonstrated using various PCM platforms, including $Ge_2Sb_2Te_5$ (GST),[23–32] $Ge_2Sb_2Se_4Te_1$ (GSST),[33] and vanadium dioxide ($VO_2$).[9,34,35] By switching between amorphous and crystalline states through optical, electrical, or thermal excitation, PCMs enable significant refractive index modulation for large-scale optical control. Therefore, the integration of PCMs with photonic platforms offers a powerful route toward compact, multifunctional, and actively reconfigurable photonic devices.[36,37] Beyond linear switching functionalities, nonlinear optical processes, such as second-harmonic generation (SHG) and third-harmonic generation (THG), provide additional capabilities for new wavelength generation and signal processing. Dielectric materials such as silicon (Si), gallium arsenide (GaAs), and lithium niobate (LN) have been widely employed for nonlinear metasurfaces due to their strong nonlinear susceptibilities and effi-cient harmonic generation.[38–42] However, their nonlinear responses are generally fixed once fabricated or have limited tunability. Various active approaches have been explored for tuning of nonlinear generations, including liquid crystals,[43–45] electrical Fermi-level tuning of graphene,[46,47] photo-induced refractive index change of polymers,[48] and PCMs.[49–52] However, simultaneously achieving large-scale resonance tuning, efficient nonlinear enhancement, and actively tunable nonlinear emission over a broad spectral range remains challenging and yet to be further explored.

In recent years, $Sb_2S_3$ has attracted significant attention for tunable nanophotonic applications due to its low optical loss in the visible and near-infrared spectral regions, together with the large refractive index contrast between its amorphous and crystalline phases. [53–59] These advantages are particularly pronounced in the telecommunication spectral range, where both phases remain nearly lossless while maintaining a stable refractive index contrast, unlike other PCMs such as GST and $VO_2$ (Figure 1(a–c)). In addition, $Sb_2S_3$ has a strong third-order nonlinear optical response, with a reported nonlinear susceptibility of $\chi^{(3)} = 4.55 \times 10^{-11}$ esu for $Sb_2S_3$ nanorods,[60] suggesting a promising platform for realising efficient tunable nonlinear metasurfaces with a large tuning spectral range.

In this work, we investigate the linear and nonlinear optical responses of monolithic and hybrid $Sb_2S_3$-based metasurfaces, demonstrating efficient resonance tuning for linear optical switching and tunable visible light generation based on THG with $Sb_2S_3$-based resonant metasurfaces, through optically induced phase-change modulation of $Sb_2S_3$. In the linear regime, we design and fabricate $Sb_2S_3$ metasurfaces supporting strong electric and magnetic dipole resonances in the telecommunication range. By optically inducing phase transition in the $Sb_2S_3$ resonators using a pump laser with a power of 110 mW corresponding to an optical energy density of $7\ \mathrm{kJ \cdot cm^{-2}}$, and ~50% crystallisation, we experimentally achieve transmission modulation depth of up to 92% near the resonance. Furthermore, by depositing a silicon layer on top and forming a hybrid $Sb_2S_3$-Si metasurface, switching power can be significantly reduced through the formation of high-Q Fano resonances based on the multipolar interfer-ence effect, which enables enhanced near-field confinement and sharper resonance linewidth, leading to a reduced switching optical energy density from 7 kJ · $cm^{-2}$ to 4.1 kJ · $cm^{-2}$. Owing to the strong resonance-dependent near-field enhancement, by pumping with a fixed broadband laser incidence, these tunable metasurfaces also exhibit dynamically tunable resonant THG emission in the visible over a ∼ 40 nm spectrial range through phase change transition. The optically-induced phase-change modulation of $Sb_2S_3$ and resonant modes provides a powerful platform for active and large-scale tuning of the resonant nonlinear emissions with metasurfaces, offering a promising route toward compact and tunable visible nonlinear light sources based on ultra-thin nonlinear metadevices.

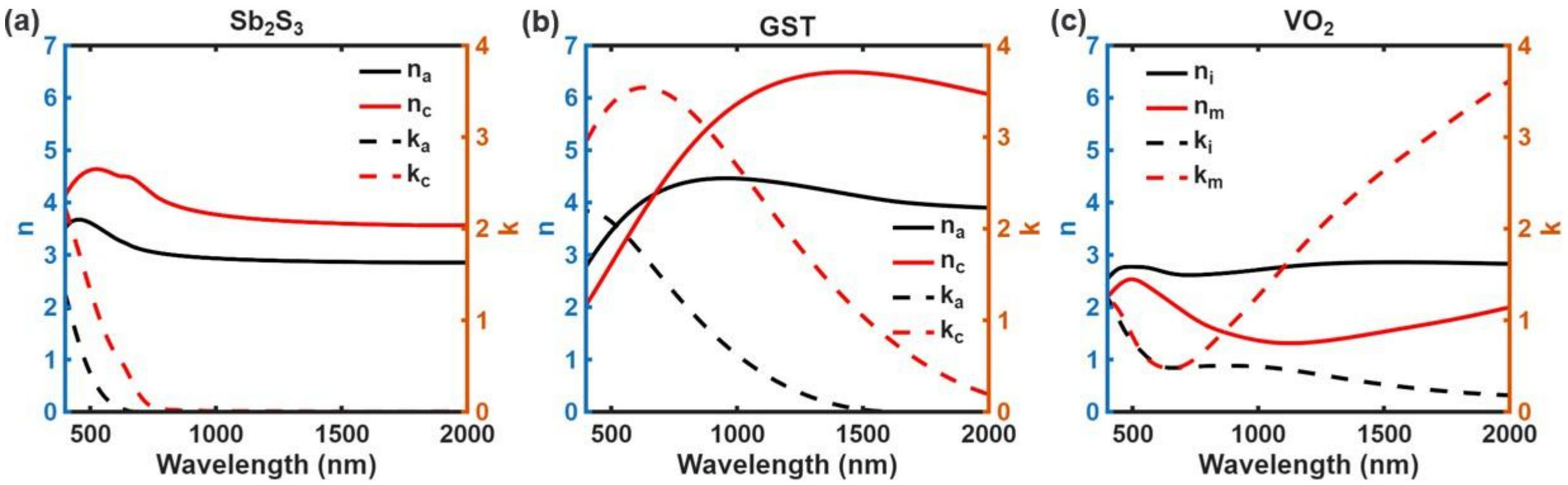

Figure 1: Complex refractive indices of three PCMs across the visible and near-infrared spectrum (400–2000 nm). Refractive index $n$ (left axis) and extinction coefficient $k$ (right axis) are shown for the amorphous and crystalline (or insulating/metallic) phases of (a) $Sb_2S_3$, (b) GST, and (c) $VO_2$.

## Results and discussion

We design and fabricate the $Sb_2S_3$-based metasurface consisting of periodically arranged $Sb_2S_3$ nanodisks on $SiO_2$ substrate, which supports electric and magnetic dipole (MD) resonances around 1500 nm.[61–63] The diameter and height of the nanodisk are set at $D$ = 650 nm and $h_1$ = 300 nm, respectively, with a metasurface period of 900 nm. By locally illuminating the metasurface with a pump laser, selected regions of $Sb_2S_3$ nanodisks undergo a phase transition from the amorphous to the polycrystalline state, enabling spatially controlled phase-change modulation. Figure 1(a) gives the refractive indices of the $Sb_2S_3$ film in both the amorphous and crystalline states measured by spectroscopic ellipsometry over a broad spectral range. In the telecommunication band, the real part of the refractive index contrast $\Delta n$ between the amorphous and crystalline phases is approximately 0.74. Such a large refractive index contrast enables substantial optical modulation required for efficient switching. For simplicity, a linear interpolation between amorphous and crystalline states was used to estimate the refractive index for intermediate crystallisation states (see Section I - Parameters of $Sb_2S_3$ material in the Supporting Information). These intermediate states of $Sb_2S_3$ are formed from the coexistence of amorphous and orthorhombic crystalline phases with different phase fractions, while a fully crystalline state corresponds to a pure orthorhombic phase.[64,65] Figure 2(ai) shows a scanning electron microscopy (SEM) image of the fabricated $Sb_2S_3$ nanodisk array, demonstrating its high structural quality. Figure 2(aii) shows an optical microscope image of the amorphous and laser-induced polycrystalline regions of the metasurface. In the microscope image, the polycrystalline regions appear darker due to stronger absorption, allowing a clear visual distinction from the optical microscope images between regions with different $Sb_2S_3$ phases. We then deposit a 94 nm-thick silicon overlayer on top of the $Sb_2S_3$ disk metasurface and form a hybrid $Sb_2S_3$/Si metasurface, as illustrated in Figure 2(b). The linear and nonlinear optical tuning with $Sb_2S_3$-based metasurfaces is illustrated in Figures 2(ci) and 2(cii), showing phase change–induced optical switching in the linear regime and visible nonlinear light generation via THG, respectively.

In our experiment, the phase transition of the $Sb_2S_3$ based metasurface was achieved through laser-induced crystallisation of $Sb_2S_3$ in the metasurface using a continuous wave optical laser beam at the wavelength of 532 nm. The spatial irradiance distribution of the excitation laser beam with a Gaussian beam profile can be described by

$$I(r) = \frac{2P}{\pi\omega_0^2} \exp\left(-\frac{2r^2}{\omega_0^2}\right) \tag{1}$$

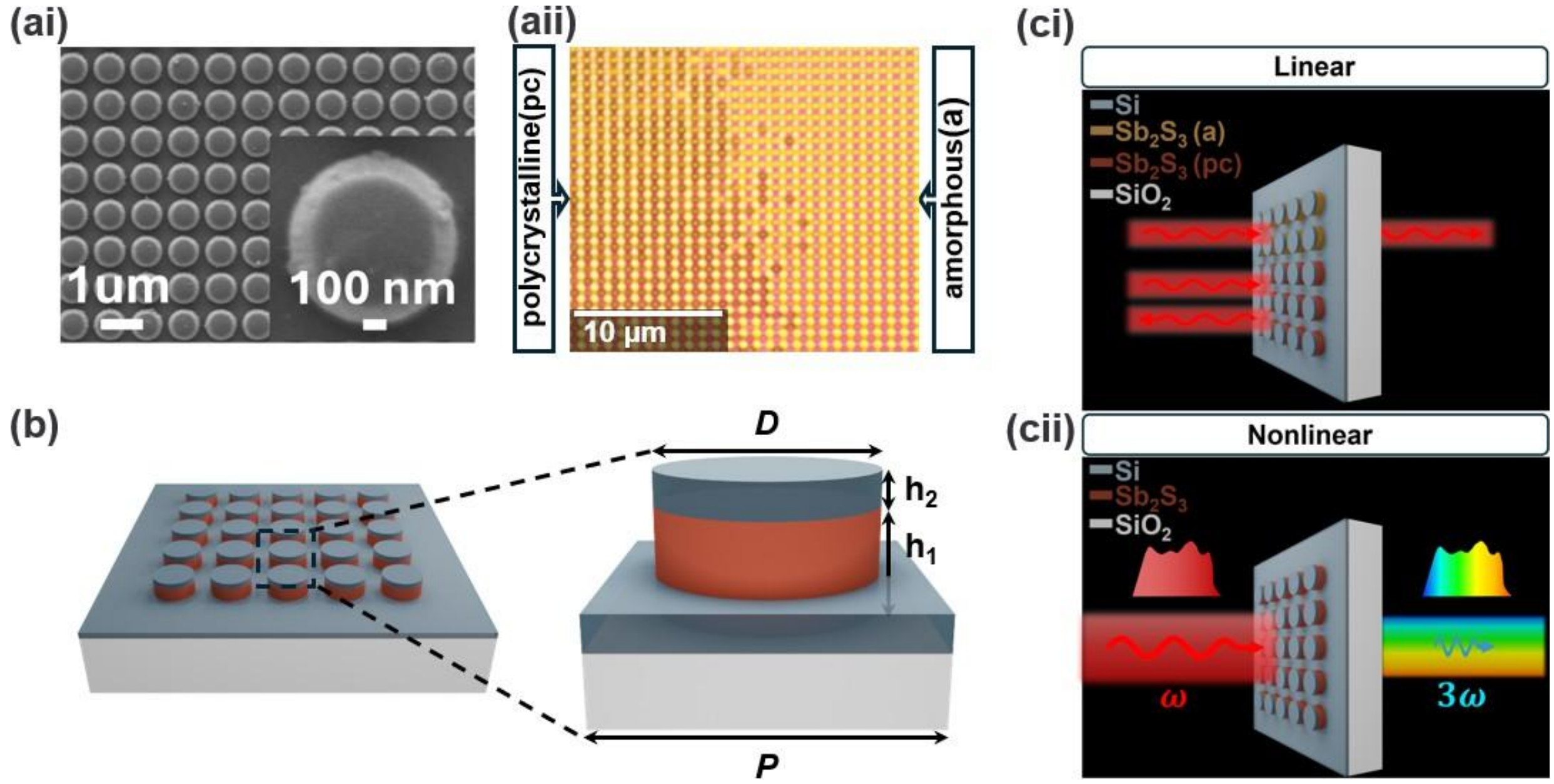

Figure 2: (a) SEM and optical microscope images of the $Sb_2S_3$ metasurface. (b) Schematic of the $Sb_2S_3$/Si hybrid metasurface and its unit cell, with diameter $D$ = 650 nm, pillar height $h_1$ = 300 nm, silicon thickness $h_2$ = 94 nm, and lattice periodicity $P$ = 900 nm. (ci, cii) Schematic illustration of phase change–induced optical switching in the linear regime and tunable visible light generation via THG.

where $\omega_0$ is the waist of the beam, $r$ is the radial distance from the centre of the beam, and $P$ is the laser power. For crystallisation, the exposure time was fixed at $t$ = 100 ms. The crystallisation threshold power was experimentally determined to be 60 mW, above which crystallisation occurs. Consequently, the crystallisation threshold energy density was calculated using Equation 1 to be $E_0 = I(r = 0) \times t = 3.8$ kJ, cm$^{-2}$. These values define the optical energy window required to drive phase transitions in $Sb_2S_3$. Reversibility can be achieved using short, high-energy optical pulses that induce rapid thermal cycling, enabling restoration of the amorphous phase and recovery of the initial optical state, as demonstrated in previous studies.[53]

We first experimentally measured the transmission spectra of the monolithic $Sb_2S_3$ disk metasurface in the amorphous and polycrystalline states (for details of the experimental setup, see Fig. S1 in the Supporting Information). Figure 3(a) shows the transmission spectra of the amorphous $Sb_2S_3$ disk metasurface under $y$-polarised light illumination for different incident angles $\theta$ when the light incidence is tilted within the $xz$ plane. Here, the incident angle $\theta$ is defined the angle between the propagation direction of the incident beam and the surface normal of the metasurface within the $xz$ plane. As can be seen, a broad resonance is observed around 1410 nm. Importantly, the resonance position remains nearly unchanged with increasing incident angle, indicating the angular robustness of the resonant mode under $y$-polarised light excitation. Subsequently, the metasurface was locally crystallised using a laser power of 110 mW and an exposure time of 100 ms, corresponding to an energy density of 7 kJ, cm$^{-2}$, to induce phase transitions from the amorphous to ~50% polycrystalline state. The corresponding transmission spectra are shown in Figure 3(b). We observe resonance broadening together with a reduction in the overall transmission after the optically induced crystallisation process. This behaviour can be attributed to possible inhomogeneous intermediate crystallisation states on the metasurface region due to the laser spot intensity variations in space, which as a result will reduce the collective effect from the periodic system and increase the dissipative losses of the system. Despite these effects, we experimentally observe a pronounced resonance shift of approximately 150 nm, driven by the increased refractive index of polycrystalline $Sb_2S_3$. As shown in Figure 3(c), the resonance wavelength redshifts from 1410 nm to 1560 nm after laser-induced crystallisation. As a result, a transmission modulation depth of up to 92%, defined as $(T_{\max} - T_{\min})/T_{\max}$, is experimentally achieved near 1560 nm. This demonstrates an efficient efficient optical ON-OFF switching through laser-induced phase transitions in the metasurface. Importantly, as the resonance remains stable over a broad angular range, efficient optical switching can be obtained across a wide range of incident angles, as shown in Figure 3(d). Such angle-insensitive switching behaviour may be beneficial for practical tunable photonic systems, enabling optical modulation beyond strictly normal-incidence operation conditions. It is worth noting that under $x$-polarised light excitation, the resonance exhibits clear spectral shifting and mode splitting with varying incident angle (more details in Fig. S2 in the Supporting Information). This angle-dependent evolution of the resonance can be used as an additional tuning degree of freedom for engineering the spectral response of the metasurface, providing possibilities for adaptable operation across different wavelength regimes. Using rigorous coupled wave analysis, [66] we numerically simulated the transmission spectra for both the amorphous and polycrystalline $Sb_2S_3$ disk metasurfaces under both *x*- and *y*-polarised light illumination. The simulated spectra are in good agreement with the experimental measurements, indicating good fabrication quality of our metasurface platform. For details of the corresponding simulated results for $Sb_2S_3$ disk metasurfaces, see Fig. S3 and Fig. S4 in the Supporting Information.

To gain a deeper insight into the resonant response of the metasurface, we perform the multipolar analysis of the $Sb_2S_3$ disk metasurface using the field within the unit cell. [67,68] Figure 4(a) shows the calculated multipolar excitations under *y*-polarised light illumination for the amorphous $Sb_2S_3$ disk metasurface. As can be seen, the resonance centred around 1410 nm is predominantly governed by the excitation of electric dipole (ED) and magnetic dipole (MD) resonances, together with a small portion of electric quadrupole (EQ), and magnetic quadrupole (MQ) contributions. By further decomposing the Cartesian ED and toroidal dipole (TD) components, as shown in the inset of Figure 4(a), we also identify an additional contribution from the TD excitation. These multipolar excitation and their interference further leads to the enhanced near-field distributions shown in Figure 4(b), where an electric field enhancement factor of 6.7 is obtained.

Integrating $Sb_2S_3$ with other dielectric materials such as Si offers a versatile platform to engineer hybrid optical resonances with enhanced Q-factors, while simultaneously enabling efficient optical tuning of the optical resonances through phase-change modulation. Si has been widely used in integrated photonics and optoelectronics, possesses a high refractive index, low absorption loss at the telecommunication wavelengths, and strong electromagnetic field confinement capability. We then investigate the complementary advantages of both $Sb_2S_3$ and Si by depositing the Si overlayer onto the $Sb_2S_3$ disk metasurface, forming a hybrid $Sb_2S_3$/Si metasurface configuration. We first numerically optimised the silicon overlayer thickness in the hybrid metasurface design to support a high-Q resonance within the 1400–1600 nm spectral range (for details, see Section III - Optimisation of Si layer in hybrid $Sb_2S_3$/Si Metasurfaces, in the Supporting Information). As a result, by depositing a 94-nm-thick Si overlayer on the $Sb_2S_3$ disk metasurface, it introduces strong resonant mode hybridisation and leads to the emergence of high-Q resonances with substantially narrowed linewidths, providing a powerful platform for enhanced light-matter interactions and efficient optical tuning (see Section IV - Linear characterisation of hybrid $Sb_2S_3$/Si metasurfaces in the Supporting Information). Figure 4(c,d) gives the calculated multipolar content for the amorphous hybrid $Sb_2S_3$/Si metasurface and the corresponding near-field distribution at the resonant position around 1464 nm. Compared with the pure $Sb_2S_3$ disk metasurface, the introduction of the Si overlayer induces additional excitation of higher-order multipolar resonances, including electric octupole (EO), magnetic quadrupole (MQ) and electric quadrupole (EQ) resonances, together with a stronger contribution from the TD excitation. The coexistence and hybridisation of these multipolar resonances give rise to the pronounced Fano interferences. As a result, strong electric near-field enhancement is achieved, with the electromagnetic field predominantly confined inside the $Sb_2S_3$/Si hybrid structure. Such highly confined field distributions are expected to expand the phase-change-driven light-matter interactions, and enable efficient large-scale tunable optical response in the hybrid metasurface system, which cannot be achieved with pure silicon or other material-based metasurfaces.

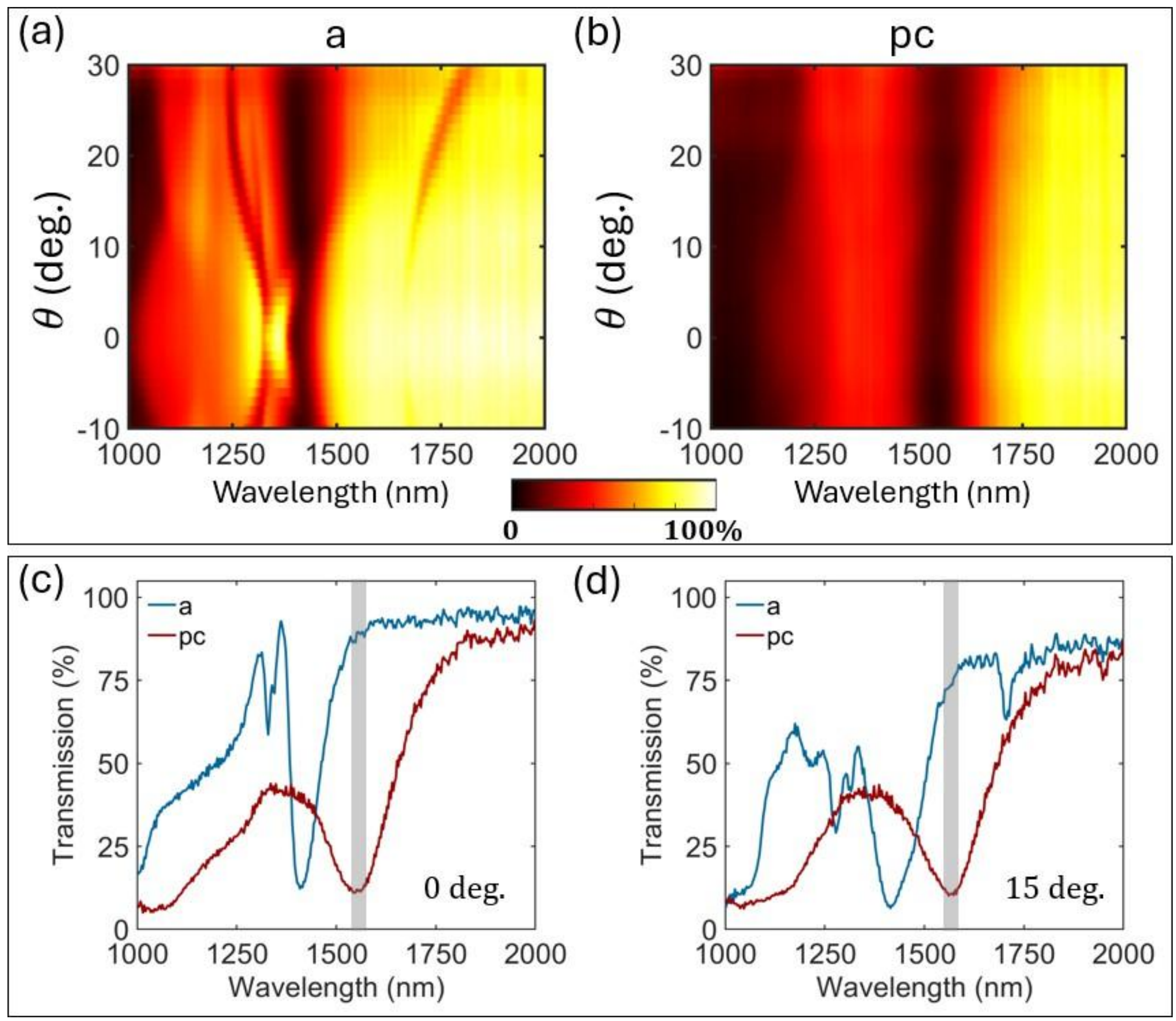

Figure 3: Experimentally measured transmission spectra of the monolithic $Sb_2S_3$ disk metasurface with different incident angles $\theta$ under $y$-polarised light illumination, for (a) amorphous and (b) polycrystalline states, respectively. (c) Experimentally measured transmission spectra for amorphous and polycrystalline states at incident angles being (c) 0 deg and (d) 15 deg.

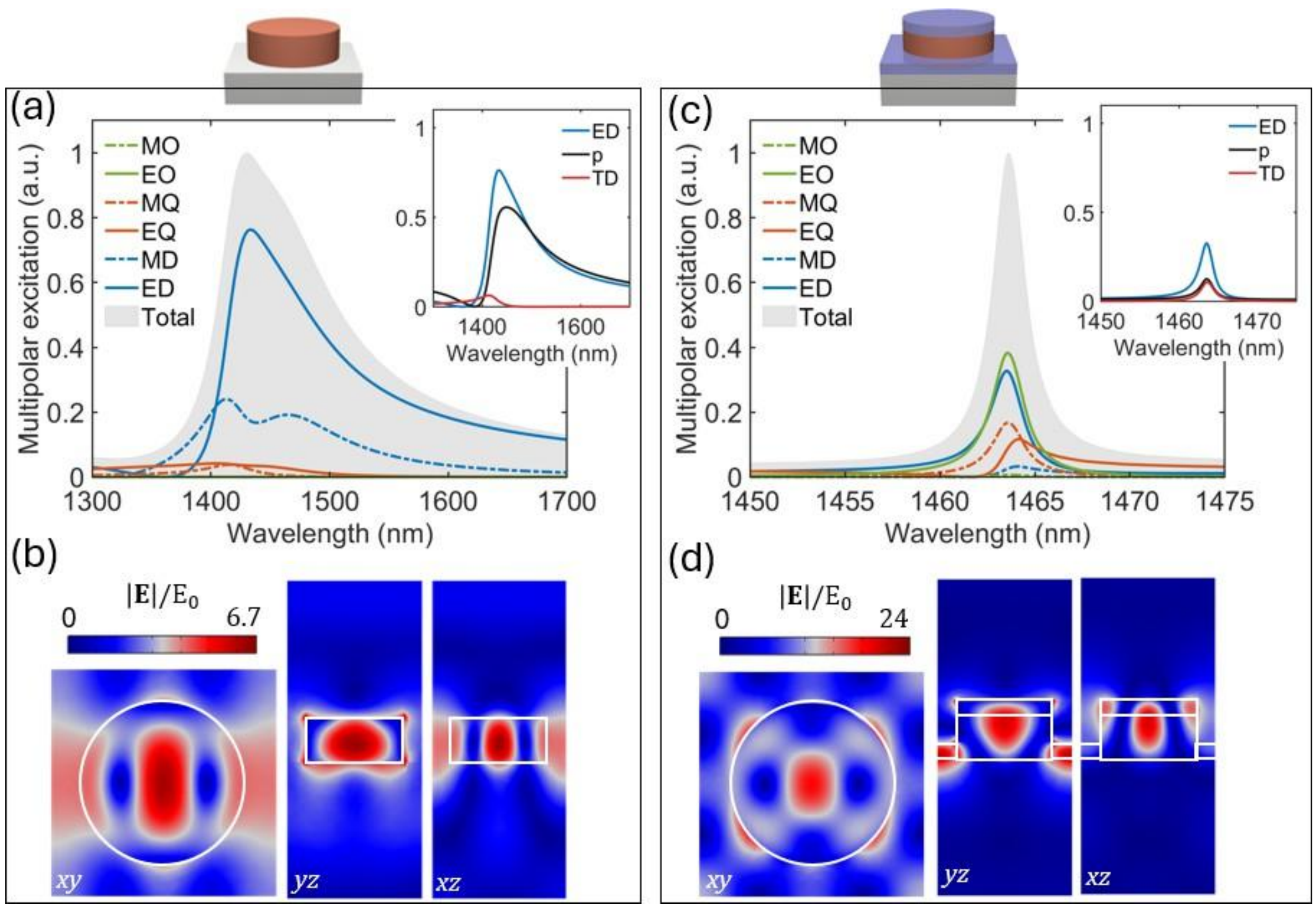

Figure 4: (a,b) Calculated linear multipolar content for the $Sb_2S_3$ disk metasurface, and the corresponding electric near-field distributions under $y$-polarised pump at 1410 nm in the vicinity of the resonant position. (c,d) Calculated linear multipolar content for the hybrid $Sb_2S_3$/Si metasurface, and the corresponding electric near-field distribution under $y$-polarised pump at 1464 nm in the vicinity of the resonant position.

We then experimentally measured the transmission spectra of the hybrid $Sb_2S_3$/Si metasurface under $y$-polarised light illumination at different incident angles for both the amorphous and polycrystalline state, as shown in Figure 5(a,b). Compared to the monolithic case in Figure 3, multiple sharp resonances emerge across the telecommunication spectral range, arising from the high-Q hybridised resonant modes supported by the coupled $Sb_2S_3$/Si metasurface system (more details in Fig. S7 in Supporting Information). Taking the Fano resonance around 1464 nm as an example, the strong multipolar mode hybridisation leads to highly confined electric near fields within the $Sb_2S_3$/Si nanostructures (see Figure 4(d)).

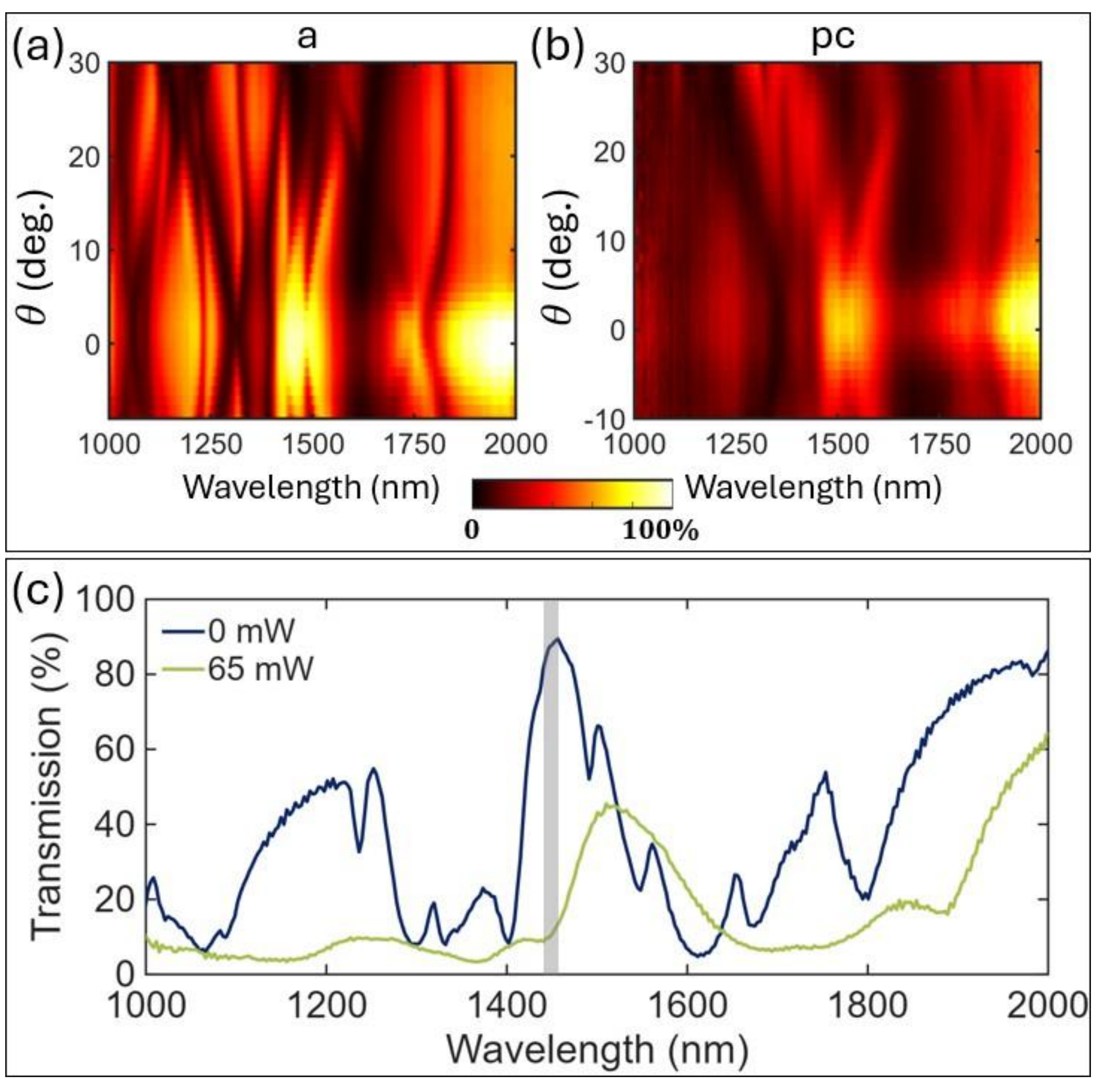

Figure 5: Experimentally measured transmission spectra of the hybrid $Sb_2S_3$/Si metasurface with different incident angles $\theta$ under $y$-polarised light illumination, for (a) amorphous and (b) polycrystalline states, respectively. (c) Experimentally measured transmission spectra for amorphous and laser-induced polycrystalline state at normal light incidence.

Such strong field localisation can greatly enhance the light-matter interaction within the phase-change material, making the resonant response highly sensitive to the refractive index variations in $Sb_2S_3$. As a result, even a relatively small phase-change modulation of $Sb_2S_3$ can produce a substantial transmission switching response. Figure 5(c) gives the transmission modulation in amorphous polycrystalline states corresponding to 0 and 65 mW pump powers. The $Sb_2S_3$/Si metasurface was exposed to green CW laser excitation with powers of 61 to 65 mW (more details in Fig. S9 in Supporting Information). As shown in Figure 5c, increasing the laser power to 65 mW induces a pronounced modulation dip around 1446 nm, resulting in an experimentally measured transmission modulation depth of approximately 90%. It is worth noting that, the hybrid $Sb_2S_3$/Si metasurface configuration significantly reduces the required switching energy compared with the monolithic $Sb_2S_3$ disk metasurface. The switching energy density decreases from 7 kJ $cm^{-2}$ to 4.1 kJ $cm^{-2}$ (110 mW vs. 65 mW laser power, respectively), using an exposure time of 100 ms and a focused beam diameter of 20 $\mu$m. This reduction in switching energy demonstrates the effectiveness of hybridised high-Q resonances and the enhanced light-matter interactions in the configuration, and this further has the potential to boost the nonlinear light-matter interactions for nonlinear optical switching and tuning. Similarly, we observe a broader resonance as compared to the simulation which is likely attributed to the non-uniform crystallisation phases across the hybrid $Sb_2S_3$/Si metasurface. For the simulated transmission spectra, see Section IV - Linear characterisation of hybrid $Sb_2S_3$/Si metasurfaces in the Supporting Information.

Beyond linear optical modulation, the strong near-field confinement and enhanced light-matter interactions supported by the $Sb_2S_3$-based metasurfaces ($Sb_2S_3$ disk metasurface and hybrid $Sb_2S_3$/Si metasurface) also provide an ideal platform for nonlinear optical generation. Due to the intrinsically large third-order nonlinear susceptibility of $Sb_2S_3$ and the enhancement of the resonant electromagnetic field, efficient third-harmonic generation is expected from the metasurface. Furthermore, phase-change-induced refractive index modulation enables active control of the resonant conditions, thus allowing dynamic switching and tuning of the nonlinear THG emission. In the following, we will investigate both numerically and experimentally the generation of THG from our $Sb_2S_3$ disk and hybrid $Sb_2S_3$/Si metasurfaces. For nonlinear optical characterisation, we use a broadband supercontinuum laser source (NKT Photonics SuperK FIU-15) to generate an infrared excitation pump beam. A longpass dichroic mirror with a cutoff wavelength of 1180 nm (DMLP1180) was used to spectrally filter the pump, and to select wavelengths between 1180 nm and 2000 nm for metasurface excitation (See Fig. S10 of the experimental setup in the Supporting Information). The pump beam was then focused onto the sample using a lens with a focal length of 7.5 cm. The resulting THG signals from the metasurface were then experimentally collected using a 20× Objective with NA=0.4, under both $y$- and $x$-polarised excitation for incident angles varying from 0° to 12°, enabling experimental investigation of the angular and polarisation dependent nonlinear optical responses of the metasurfaces. Figure 6(a-d) shows the experimentally measured THG emission from the monolithic $Sb_2S_3$ disk metasurface under both $y$- and $x$-polarised pump excitation. As can be seen, strong THG emission is observed from the amorphous $Sb_2S_3$ disk metasurface in the vicinity of the resonance excited by pump wavelengths around 1410 nm, consistent with the linear transmission spectra shown in Figure 3 and the corresponding multipolar content in Figure 4(a). The enhanced THG response originates from the resonantly enhanced electromagnetic near fields supported by the coupled ED and MD-dominated modes within the metasurface. Importantly, due to the phase-change-induced modulation of the refractive index of $Sb_2S_3$, the THG emission can be spectrally tuned by laser-induced crystallisation. Following the transition from the amorphous to the polycrystalline state, the THG emission undergoes a pronounced spectral redshift from visible blue to green. This behaviour demonstrates the capability of the phase-change metasurface for dynamically tunable nonlinear wavelength conversion and tunable visible light generation. Furthermore, due to the polarisation-dependent angular response of the resonance, we experimentally observe angle-dependent THG emission under different pump excitation polarisations. Under $y$-polarised pump excitation, the THG emission remains spectrally stable over a broad range of incident angles, suggesting the angular robustness of the underlying resonance as mentioned in the linear responses shown in Figure 3. In contrast, under *x*-polarised pump excitation, the THG wavelength gradually redshifts with increasing incident angle, enabling continuous visible colour tuning from blue to cyan through angular control of the pump beam. We further numerically calculated the THG emission power assuming a normal incidence plane wave with a power density 1 $GW/cm^2$. Figure 6(e) gives the calculated THG emission power for amorphous and 50% polycrystalline states of the $Sb_2S_3$ disk metasurfaces (see Section VI - Numerical methods for nonlinear simulation for details of the simulation method in the Supporting Information). By comparing the simulation and experimental results, we observe a reduction in the experimentally measured THG emission for the polycrystalline case. The experimentally observed reduction in THG efficiency is likely associated with the spatially inhomogeneous intermediate crystallisation states across the metasurface, arising from the non-uniform intensity distribution of the laser beam. Such partial and non-uniform phase transitions can disrupt the collective resonant response of the periodic system, leading to resonance broadening, reduced field confinement , and enhanced optical dissipation. Figure 6(f) gives the calculated electric near-field distribution at the THG wavelength when pumping at the resonance position 1464 nm for the hybrid $Sb_2S_3$/Si metasurface in its amorphous state. The spatial field profile of the THG signal closely resembles the near-field distribution observed at the fundamental pump resonance (see Figure 4(d)), indicating that the enhanced THG emission predominantly originates from resonance-enhanced electromagnetic confinement at the pump wavelength.

Building on the strong linear optical modulation and enhanced near-field confinement observed in the hybrid $Sb_2S_3$/Si metasurface, similarly, we further experimentally characterise its nonlinear THG response. As demonstrated in the linear regime, the hybrid metasurface supports high-Q Fano resonances arising from strong multipolar mode hybridisation, leading to highly localised electromagnetic fields within the $Sb_2S_3$/Si nanostructures. Figure 7(a-f) presents the experimentally measured THG emission from the hybrid $Sb_2S_3$/Si metasurface under different excitation conditions, and Figure 7(g, h) give the calcualted THG emission power for the amorphous, 25% and 50% crystalline states, as well as the calculated electric near-field distributions at the THG wavelength under resonant pumping for the amorphous hybrid $Sb_2S_3$/Si metasurface. Compared with the monolithic $Sb_2S_3$ metasurface, the hybrid configuration exhibits enhanced THG emission intensity owing to the sharper resonance linewidths and increased near-field enhancement associated with the high-Q hybrid resonances. Together with the polarisation- and angle-dependent resonant characteristics (Figure 5, and Section IV - Linear characterisation of hybrid $Sb_2S_3$/Si metasurfaces in the Supporting Information), the hybrid $Sb_2S_3$/Si metasurface provides a versatile platform for dynamically controllable visible nonlinear light sources and reconfigurable nonlinear nanophotonic devices. It is worth noting that the observed THG emission in the hybrid metasurface originates from the combined nonlinear responses of both the $Sb_2S_3$ phase-change material and the silicon overlayer. While $Sb_2S_3$ possesses an intrinsically strong third-order nonlinear susceptibility, silicon is widely known for its highly efficient third-order nonlinear optical response and strong THG generation capability. Under resonant excitation, the electromagnetic fields are strongly confined within both the $Sb_2S_3$ and silicon regions of the hybrid structure, allowing simultaneous nonlinear contributions from the two materials. More broadly, the hybrid integration approach demonstrated here can be generalised to other nonlinear photonic platforms, including lithium niobate (LN) and GaAs-based devices, to achieve efficient tunable nonlinear optical devices with a large tuning spectral range. Although LN, Si, GaAs material-based metasurfaces can intrinsically support strong second-order or third-order nonlinear processes through large nonlinear susceptibilities and resonantly enhanced electromagnetic fields, their nonlinear responses are generally fixed once fabricated and lack efficient post-fabrication tunability with a large tuning spectral range. By integrating these nonlinear materials with phase-change $Sb_2S_3$, the strong nonlinear emission generated within the host resonant structure can be dynamically controlled through phase-change-induced modulation of the hybrid resonances. Such integration combines the efficient nonlinear conversion capability of conventional nonlinear materials with the large optical tunability of $Sb_2S_3$, thereby providing a promising pathway toward actively programmable nonlinear metasurfaces and compact tunable light generators based on nonlinear optics.

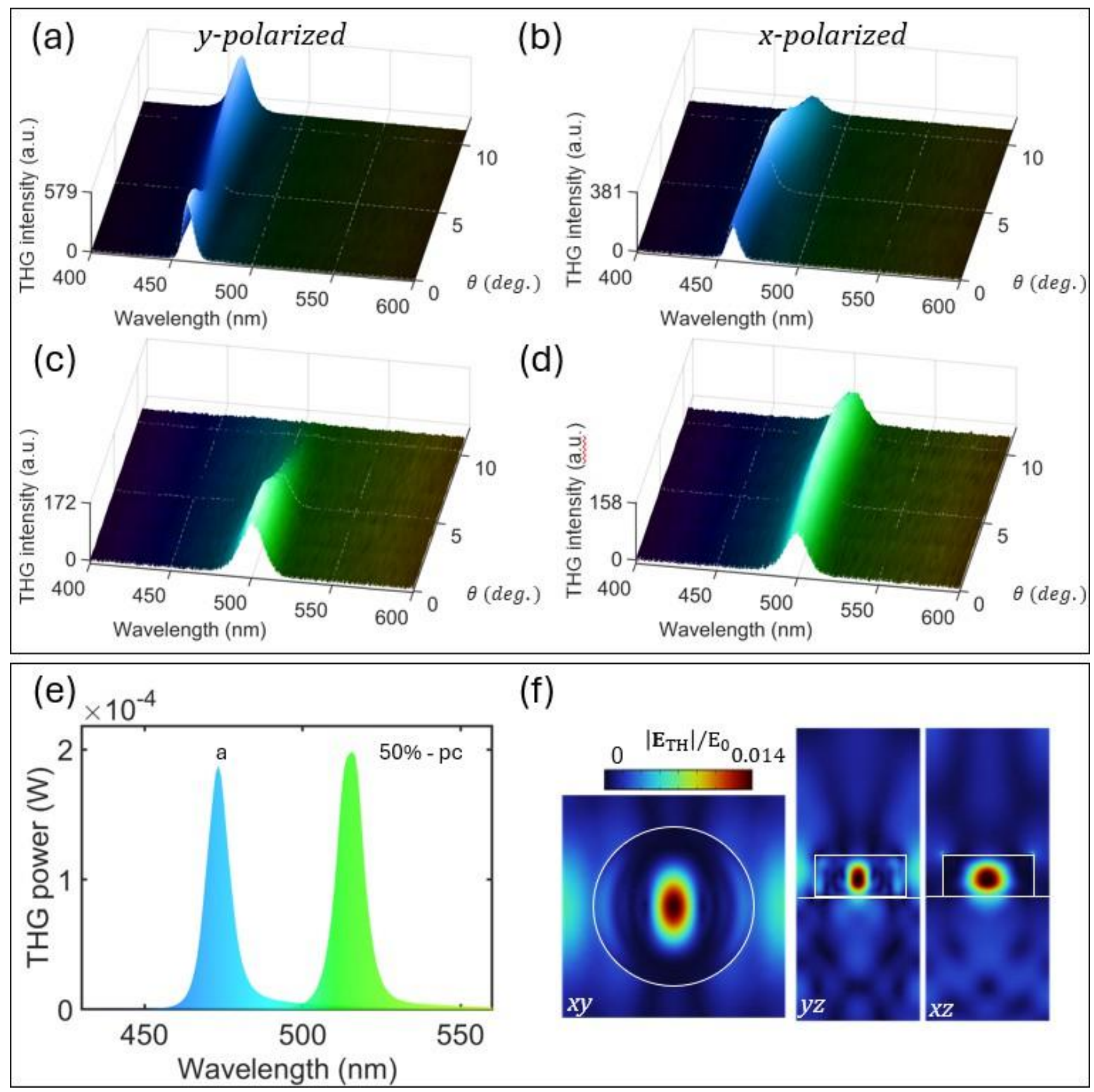

Figure 6: Experimentally measured THG emission intensities of monolithic $Sb_2S_3$ disk metasurfaces under *y*-polarised (a, c) and *x*-polarised (b, d) supercontinuum illumination for the amorphous (a, b) and polycrystalline (c, d) states at different incident angles. (e) Simulated THG emission power under normal-incidence pumping with a power density of 1 GW/cm$^2$ for the amorphous and 50%-crystalline states. (f) Calculated electric near-field distributions at the THG wavelength under resonant pumping for the amorphous $Sb_2S_3$ disk metasurface.

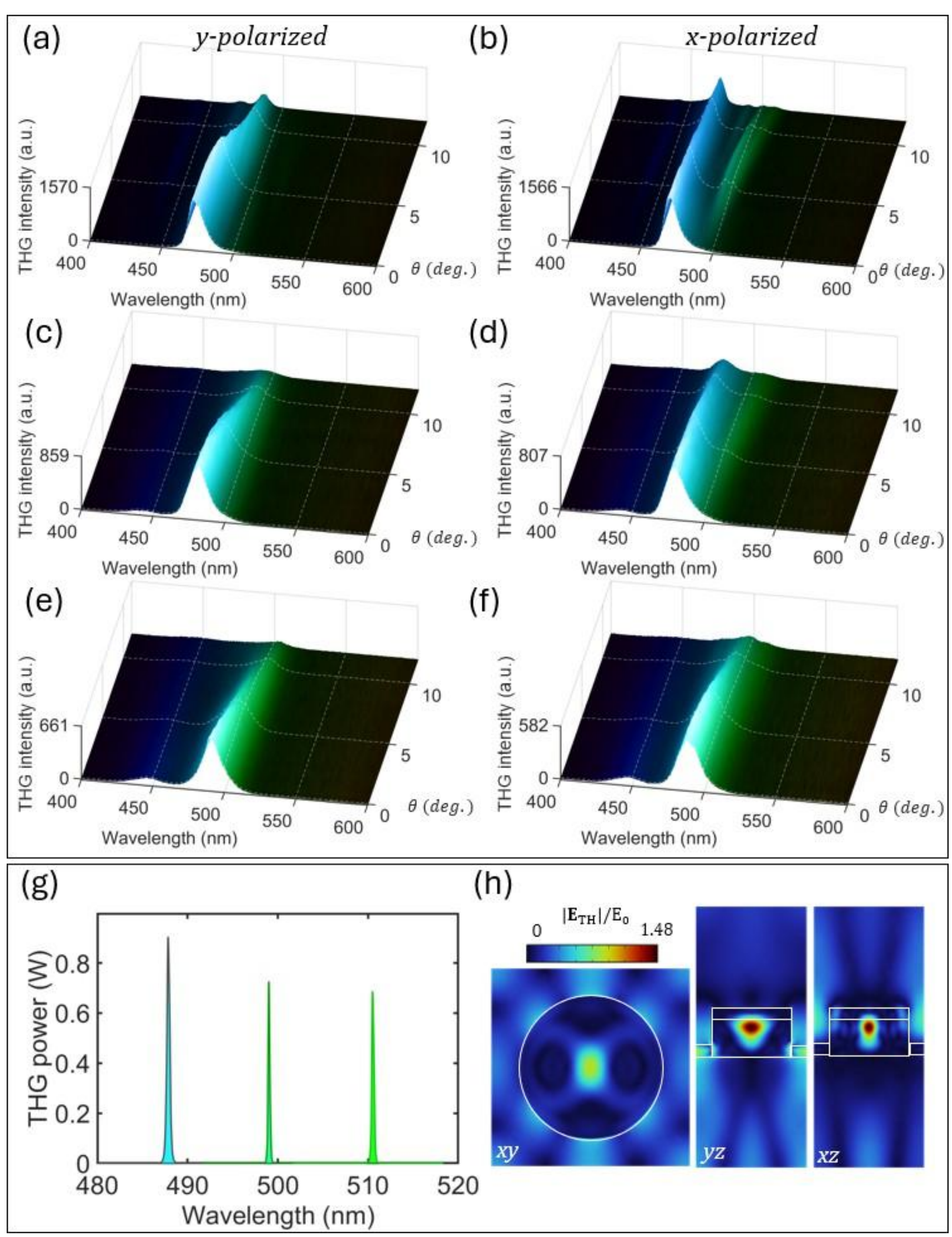

Figure 7: Experimentally measured THG emission intensities of hybrid $Sb_2S_3$/Si metasurfaces under *y*-polarised (a, c, e) and *x*-polarised (b, d, f) supercontinuum illumination for the amorphous (a, b) and approximately 25% (c, d) and 50% crystalline (e, f) states states at different incident angles. (e) Simulated THG emission power under normal incidence pumping with a power density of 1 GW/cm$^2$ for the amorphous, 25% and 50% crystalline states. (f) Calculated electric near-field distributions at the THG wavelength under resonant pumping for the amorphous hybrid $Sb_2S_3$/Si metasurface.

## Conclusion

In conclusion, we have experimentally demonstrated actively tunable linear and nonlinear optical responses in both monolithic $Sb_2S_3$ and hybrid $Sb_2S_3$/Si metasurfaces operating in the telecommunication spectral range. In the linear regime, the monolithic $Sb_2S_3$ metasurface supports magnetic dipole-dominated resonances with strong angular robustness, enabling experimentally measured transmission modulation depths of up to 92% through laser-induced phase transitions. By further integrating a silicon overlayer, high-Q hybrid Fano resonances with strongly enhanced electromagnetic field confinement are achieved, substantially reducing the switching energy density from 7 kJ, $cm^{-2}$ to 4.1 kJ, $cm^{-2}$ while maintaining efficient optical switching performance. In the nonlinear regime, strong resonance-enhanced third-harmonic generation is experimentally observed from both the monolithic $Sb_2S_3$ disk metasurface and hybrid $Sb_2S_3$/Si metasurface. Owing to the combined nonlinear contributions from $Sb_2S_3$ and Si, together with the enhanced near-field localisation supported by the hybrid resonances, we obtain $\sim$ 3 times stronger THG emission in the visible spectral range as compared to the monolithic case. Moreover, dynamic tunability of THG emission is achieved through phase-change modulation, polarisation control, and angle-dependent resonance engineering. The demonstrated nonlinear emission tuning from blue to green suggests that the phase-change metasurface platform is a promising option for actively controllable visible light generation. More broadly, the hybrid integration strategy introduced here provides a versatile pathway for combining phase-change materials with highly nonlinear photonic platforms such as silicon, lithium niobate, GaAs, etc, to realise dynamically reconfigurable nonlinear metasurfaces. The ability to simultaneously engineer resonance hybridisation, nonlinear light–matter interactions, and active optical tunability opens new opportunities for compact low-power optical switches, programmable nonlinear emitters, and active reconfigurable nanophotonic devices.

## Acknowledgement

This research work has been kindly supported by the Royal Society and the Wolfson Foundation. The authors appreciate the use of the NTU Imaging Suite, NTU High Performance Computing cluster Avicenna, and NTU Medical Technologies Innovation Facility (MTIF). G.S. acknowledges support from the Nottingham BBSRC. M.Q. acknowledges support from the National Natural Science Foundation of China (Grant No. 12364049) and the Natural Science Foundation of Jiangxi Province (Grant No. 20242BAB25041).

## Supporting Information Available

Experimental procedures and additional characterisation data supporting the results of this study are provided in the Supporting Information.